\documentclass{emulateapj}

\shorttitle{Dust in Perseus}
\shortauthors{Schnee, S. et al.}

\begin{document}

\newcommand{\cso}{C$^{17}$O}
\newcommand{\ceo}{C$^{18}$O}
\newcommand{\cs}{C$^{34}$S}
\newcommand{\dcop}{DCO$^+$}
\newcommand{\nthp}{N$_2$H$^+$}
\newcommand{\ntdp}{N$_2$D$^+$}
\newcommand{\kms}{km s$^{-1}$}

\title{Dust Emission from the Perseus Molecular Cloud} 
\author{S. Schnee\altaffilmark{1}, J. Li\altaffilmark{2}, A. A. Goodman\altaffilmark{2} \& A. I. Sargent\altaffilmark{1}}

\email{schnee@astro.caltech.edu}

\altaffiltext{1}{Division of Physics, Mathematics and Astronomy, 
       California Institute of Technology, 770 South Wilson Avenue,
       Pasadena, CA 91125}
\altaffiltext{2}{Harvard-Smithsonian Center for Astrophysics, 60 Garden
       Street, Cambridge, MA 02138}

\begin{abstract}
Using far-infrared emission maps taken by IRAS and Spitzer and a
near-infrared extinction map derived from 2MASS data, we have made
dust temperature and column density maps of the Perseus molecular
cloud.  We show that the emission from transiently heated very small
grains and the big grain dust emissivity vary as a function of
extinction and dust temperature, with higher dust emissivities for
colder grains.  This variable emissivity can not be explained by
temperature gradients along the line of sight or by noise in the
emission maps, but is consistent with grain growth in the higher
density and lower temperature regions.  By accounting for the
variations in the dust emissivity and VSG emission, we are able to map
the temperature and column density of a nearby molecular cloud with
better accuracy than has previously been possible.
\end{abstract}


\section{Introduction}

Far-infrared and sub-millimeter continuum radiation from interstellar
clouds is typically attributed mostly to ``dust.''  In truth, however,
even the contributions from ``dust'' come from multiple populations of
grain types, which are each at a mixture of temperatures.  Along any
given line of sight, there may be random or systematic variations both
in the admixture of grain types and in their temperature, and those
variations may or may not be correlated.  Historically, researchers
have used more and more complicated procedures to account for dust
emission and estimate column density.

The simplest way to infer column density from ``thermal'' dust
emission is to assume a single grain type, radiating with a single
efficiency (opacity), at a single temperature.  In that case, flux at
any wavelength can be converted to a column density, assuming a dust
temperature.  The next-most-sophisticated approach is to use
observations at two wavelengths, and to infer a color temperature, by
assuming that the dust emission comes from a modified blackbody, which
has a wavelength-dependent emissivity one takes as given
\citep[e.g.,][]{Wood94}.

With observations at three wavelengths, it should be possible to also
constrain the dust emissivity, assuming its functional dependence on
wavelength is known and is described by just one parameter
($\kappa_\lambda \propto \lambda^{-\beta}$).  As \citet{Schnee07} have
shown, however, a three-wavelength constraint on $T_d$, $N_d$ and
$\beta$ independently is only possible for the case of very (to date
unrealistically) high signal-to-noise data.  Nevertheless, with three
wavelengths of very high S/N, or measurements at four or more
wavelengths, one could constrain the temperature, column density and
emissivity independently -- if only all the mixtures of grains along
lines of sight were similar, and the temperature distribution along
the line of sight were isothermal.

In an earlier paper, \citep[hereafter SBG]{Schnee06}, we showed that
line-of-sight variations in the dust temperature account for nearly
all the scatter in a plot of dust-emission vs. dust-extinction derived
column density in the Perseus molecular cloud.  In \citet[hereafter
  SRGL]{Schnee05}, we found that Very Small Grains (VSG's) are likely
to account for a large fraction ($> 50$\%) of the dust emission at
wavelengths shorter that 100 \micron.  If that fraction is relatively
constant, then one can say that Big Grains (BG's) are a good proxy for
column density, modulo the uncertainty imposed by the line-of-sight
variations in the dust temperature.  However, if the VSG contribution
is not constant, but depends on either $N_d$ or $T_d$, then the
situation is further confounded.  In this paper, we explore all of
these possibilities, and we find, in agreement with earlier works
\citep[e.g.,][]{Stepnik03}, that the VSG contribution does seem to
vary systematically with $T_d$ and $N_d$, and that it is important to
account for this effect in estimates of dust column density and dust
temperature.

\section{Observations}

SRGL showed that far-infrared emission maps can be used to
determine the dust temperature and column density in molecular clouds
once the conversion between FIR optical depth and $V$-band extinction
has been determined.  This analysis took advantage of the IRIS
(Improved Reprocessing of the IRAS Survey) maps at 60 and 100
\micron\ \citep{Miville05} and near-infrared extinction map created
from 2MASS data as part of the COMPLETE (COordinated Molecular Probe
Line Extinction and Thermal Emission) Survey of Star Forming Regions
\citep{Ridge06}.  Subsequently, SBG demonstrated that the column
density determined from pairs of FIR emission maps should become more
accurate and less biased when the emission is measured at longer
wavelengths due to the decreased influence of line-of-sight
temperature variations.

With the recent release of 70 and 160 \micron\ maps of Perseus taken
with the Spitzer Space Telescope \citep{Evans03, Rebull07}, we can
improve upon the SRGL analysis.  In this short paper, we combine the
100 \micron\ IRIS and 160 \micron\ Spitzer maps to derive dust
temperature and column density maps that are not strongly biased by
the emission from transiently heated VSG's.  In addition we are able
to estimate the variation of the VSG emission and BG emissivity with
$T_d$ and $A_V$.

\subsection{Spitzer} \label{SPITZER}

We used the 70 and 160 \micron\ emission maps of the Perseus molecular
cloud from the Spitzer Science Archive that were taken as part of the
Spitzer Legacy Program ``From Molecular Cores to Planet-forming
Disks'' (c2d) \citep{Evans03, Rebull07}.  Map resolutions are
18\arcsec\ and 40\arcsec\ at 70 and 160 \micron, respectively.  The
c2d delivery includes error maps, which account for systematic and
random errors.  The overall gain calibration uncertainty is $\sim$5\%
at 70 \micron\ \citep{Gordon07} and $\sim$12\% at 160 \micron\
\citep{Stansberry07}.  The c2d error maps show that the random error
for maps smoothed to 5\arcmin\ resolution (the resolution of the
extinction map) is $<$1\% at both 70 and 160 \micron.

We applied color corrections to these archival fluxes since they were
determined assuming that the emission is from a blackbody at 10,000 K
while we have assumed that the emission comes from a modified
blackbody with emissivity spectral index $\beta = 2$ \citep{Draine84}
and equilibrium temperature to be determined, but certainly less than
10,000 K \citep{Schlegel98}.  The color correction, which is combined
with the dust temperature calculation, is futher described in
Sect.~\ref{TD}.

\subsection{IRAS/IRIS} \label{IRIS}

IRIS flux density maps at 60 and 100 \micron, in contrast with earlier
releases of the IRAS all-sky maps, have been corrected for the effects
of zodiacal dust and striping, and have the proper gain and offset
calibration \citep{Miville05}.  The IRIS maps of Perseus have a
resolution of approximately 4\farcm 3 (though we smooth them to the
5\arcmin\ resolution of our NIR extinction map for all analyses).  The
noise in the IRAS maps is approximately 0.03 MJy sr$^{-1}$ and 0.06
MJy sr$^{-1}$ at 60 and 100 \micron\ \citep{Miville05}, which is
$\lesssim$1\% of the flux in the portion of Perseus included in the
c2d survey.  Like the Spitzer maps, these were color-corrected by us
(following the IRAS Explanatory Supplement VI C.3).  For consistency,
we consider only the region of Perseus covered by the c2d MIPS maps at
70 and 160 \micron\ in this analysis.

\subsection{NIR Extinction Map} \label{2MASS}

An extinction map of Perseus based on the Two Micron All Sky Survey
(2MASS) point source catalog was created as part of the COMPLETE
survey \citep{Ridge06} using the NICER algorithm, devised by
\citet{Lombardi01}.  This algorithm uses the near-infrared color excess 
of background stars to estimate the column density of foreground dust.
When directly comparing NIR and FIR-derived column densities (e.g.,
Figs.~\ref{TAUAVTEMP}, \ref{AVSCATTER} and \ref{PLOTHIST}), all maps
have been smoothed to a common resolution of 5\arcmin, since that is
the resolution of the NICER extinction map of Perseus.

\section{Analysis} \label{ANALYSIS}

The flux density emitted by dust in thermal equilibrium is given by
\begin{equation} \label{DUSTFLUX}
S_{\nu} = \Omega B_{\nu}(T_d) N_d \alpha \nu^{\beta},
\end{equation}
where
\begin{equation} \label{BLACKBODY}
B_\nu(T_d) = \frac{2h\nu^3}{c^2} \frac{1}{\exp(h\nu /kT_d)-1}
\end{equation}

In Equation \ref{DUSTFLUX}, $S_{\nu}$ is the observed flux density at
frequency $\nu$; $\Omega$ is the solid angle of the beam;
$B_{\nu}(T_d)$ is the blackbody emission from the dust at temperature
$T_d$; $N_d$ is the column density of dust, $\alpha$ is a constant
that relates the flux density to the optical depth of the dust at
frequency $\nu$, and $\beta$ is the emissivity spectral index of the
dust, which we assume to be 2 throughout this paper \citep{Draine84}.
In SRGL we showed that $\beta = 2$ is a reasonable value for the cold
dust in Perseus.

Equation \ref{DUSTFLUX} is only true for BG's in thermal equilibrium,
and does not describe the emission from stochastically heated VSG's,
which cannot be characterized by an equilibrium temperature
\citep{Draine01}.  The VSG's are expected to be a significant
component of the observed emission in the IRAS 60 \micron\ and Spitzer
70 \micron\ bands, but are not expected to contribute much in the IRAS
100 \micron\ and Spitzer 160 \micron\ bands \citep{Li01}.  If we were
to ignore the VSG emission, we would systematically overestimate the
dust temperature and underestimate the column density in Perseus.  We
therefore account for it using a procedure that improves upon that
adopted by SRGL, as described below.

\subsection{Dust Temperature} \label{TD}

The fluxes quoted in the IRAS and Spitzer emission maps are derived
from the total energy detected convolved with the wavelength-dependent
instrumental response and an assumed SED of the emitting source.  The
quoted flux density at wavelength $\lambda_0$ is related to the
intrinsic flux density by:

\begin{equation} \label{FQUOTE}
F_{\lambda_0}^{quoted} = \frac{\int (F_\lambda^{intrinsic}/
  F_{\lambda_0}^{intrinsic}) \lambda R_\lambda d\lambda} {\int
  (F_\lambda^{default} / F_{\lambda_0}^{default}) \lambda R_\lambda
  d\lambda} F_{\lambda_0}^{intrinsic}
\end{equation}
\citep{Stansberry07} where $F_\lambda^{intrinsic}$ is the true source
SED, $F_\lambda^{default}$ is the default SED and $R_\lambda$ is the
instrumental response function (the sensivity of the detector as a
function of wavelength).  The default SED for the Spitzer data is that
of a 10,000 K blackbody, while the default SED for the IRIS data is
$F_\lambda \propto 1/ \lambda$.  We assume that the intrinsic SED is a
modified blackbody with $\beta$ = 2 \citep{Draine84}, so the color
correction is temperature dependent.

To derive the color correction and dust temperature, we create a
lookup table of the intrinsic flux density at $\lambda = $ 60, 70, 100
and 160 \micron\ for dust at temperatures between 5 and 100 K, at 0.1
K intervals.  We then calculate what flux density would have been
quoted at those wavelengths using Eq.~\ref{FQUOTE}.  Using this table,
we are able to derive the color-corrected dust temperature from the
ratio of the {\it quoted} flux densities, making the assumption that
the dust emission is optically thin and isothermal along each line of
sight.  The IRAS 100 \micron\ and Spitzer 160 \micron\ maps are not
significantly corrupted by VSG emission, so we derive a dust
temperature map using our lookup table and the quoted
$F_{100}/F_{160}$ after smoothing the emission maps to a common
(5\arcmin) resolution.

The IRAS 60 and Spitzer 70 \micron\ flux densities come from a
combination of BG and VSG emission, but Equations \ref{DUSTFLUX} and
\ref{FQUOTE} are only accurate for BG emission.  In SRGL we removed the 
VSG emission by assuming that a constant fraction of all 60 \micron\
emission can be attributed to VSG's.  This proportion was found by
minimizing the difference between the temperature maps derived from
IRIS data and the dust temperature map derived by \citet{Schlegel98},
which was derived from DIRBE 100 and 240 \micron\ emission (and
therefore free from VSG contamination).

However, the proportion of 60 and 70 \micron\ flux emitted by VSG's is
expected to vary with environment.  In regions with high extinction,
the VSG's will be shielded from the ISRF and will therefore not be
effectively heated.  In cold, dense regions dust grains may stick
together, removing the VSG population.  For instance,
\citet{Stepnik03} have shown that the VSG emission is greatly reduced
in a dense filament in Taurus where $A_V > 2.1$ and found a
corresponding increase in the dust emissivity.  A similar transition
region of altered dust composition has been reported using IRAS 60 and
100 \micron\ emission maps \citep[e.g.,][]{Laureijs91, Laureijs89,
  Boulanger90}.

Here we take advantage of the higher resolution of the IRAS 100 and
Spizter 160 \micron\ images, as compared with the DIRBE maps, to
remove the emission from VSG's accounting for spatial variations in
their contribution.  We divide our 100/160 \micron\ derived
temperature maps into ten equally populated temperature and extinction
subsets (five temperature subsets for $A_V$ less than the median
NIR-derived extinction and five temperature subsets for higher $A_V$)
and calculate the VSG contribution to the 60 and 70 \micron\ emission
independently for each bin using the dust temperature derived from the
100/160 \micron\ pair of emission maps.  In each bin, we multiply the
60 and 70 \micron\ emission (smoothed to 5\arcmin\ resolution) by the
scaling factors $f_{60}$ and $f_{70}$ which are chosen such that the
temperatures derived from $f_{60}F_{60}/F_{100}$ and
$f_{70}F_{70}/F_{160}$ are best matched to the dust temperature
derived from $F_{100}/F_{160}$.  The comparatively low resolution of
the DIRBE-derived temperature map prevented us from performing this
more sophisticated correction in SRGL.  The scale factors ($f_{60}$
and $f_{70}$) used to account for the VSG emission are shown in
Fig.~\ref{BGCONVERSION}.  Our assumptions that dust is isothermal
along each line of sight and that the dust emissivity (characterized
by $\alpha$ and $\beta$ in Eq.~\ref{DUSTFLUX}) is constant introduce
uncertainties into our calculations.  The effect of assuming
isothermality is covered in SBG, and we discuss the effects of
variable dust emissivity in Section \ref{VARDUST}.

\subsection{Column Density: Constant Dust Emissivity} \label{NH}

Given the flux density at one wavelength and the dust temperature, the 
optical depth is given by:

\begin{equation} \label{TAUEQ}
\tau_\lambda = \frac{S_\lambda}{B_\lambda (T_d)}
\end{equation}
where $B_\lambda (T_d)$ is the Planck function and $S_\lambda$ is the
observed flux density at wavelength $\lambda$.

\begin{figure}
\epsscale{1.0}
\plotone{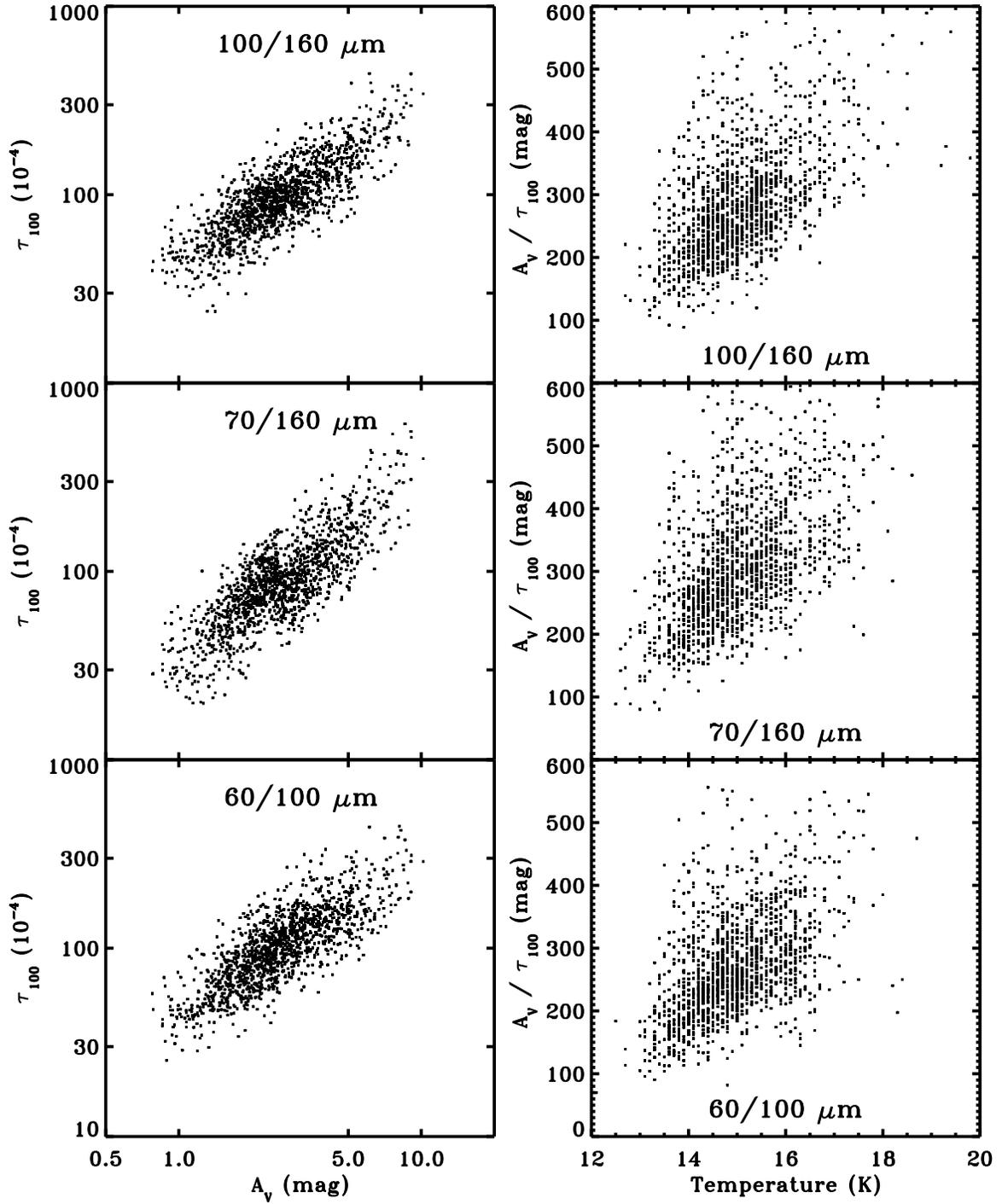}
\caption{({\it left}) The 100 \micron\ optical depth plotted against
the column density (expressed in terms of $V$-band extinction) derived
from the 2MASS map.  ({\it right}) The ratio of the NIR-derived column
density divided by the FIR-derived optical depth (at 100 \micron)
plotted against the FIR-derived dust temperature. \label{TAUAVTEMP}}
\end{figure}

The optical depth at wavelength $\lambda$ can be converted to the $V$-band
extinction (A$_V$) using the equation

\begin{equation} \label{AVEQ}
A_V = X_{\lambda} \tau_\lambda
\end{equation}
where $X_\lambda$ is the scale factor, depending on the wavelength
($\lambda$) of the emission map, that relates the thermal emission
properties of the dust grains to their $V$-band absorption properties
(Eq.~5 from SRGL).  We solve for $X_\lambda$ for each pair of emission
maps by minimizing the difference between the emission-derived column
density (smoothed to the resolution of the extinction map) and the
2MASS absorption-derived column density, as in SRGL and SBG.  For the
60/100 and 100/160 pair of emission maps we derive the optical depth
and conversion factor ($\tau_{100}$ and $X_{100}$) from the 100
\micron\ emission map.  For the 70/160 pair of emission maps we derive
$\tau_{160}$ and $X_{160}$, and scale both values in Figures
\ref{BGCONVERSION} and \ref{TAUAVTEMP} to the expected values at 100
\micron\ assuming $\beta = 2$ using the scale factor
$(\lambda_{100}/\lambda_{160})^\beta = 0.39$.

The optimization sets the overall calibration for the FIR data, but
does not force the slope of NIR-derived column density plotted against
FIR-derived column density to be exactly unity, though it will be
close to this value.  Variations in temperature along each line of
sight will affect the emission-derived $A_V$ at high column density
more than at low column density, and the extinction-derived $A_V$ is
unaffected by temperature fluctuations.  This difference in the way
the dust temperature is coupled to the derived column density, along
with possible variations in the FIR dust emissivity, can change the
slope of the emission vs. absorption-derived column density to be
slightly different from unity.  Note that Eq.~\ref{AVEQ} assumes that
the dust emission and absorption properties are constant.  The linear
relationship between $\tau_{FIR}$ and $A_V$ is shown in
Fig.~\ref{TAUAVTEMP}.

\subsection{Column Density: Variable Dust Emissivity} \label{VARDUST}

Just as the emission from VSG's is expected to vary with environment,
so should the dust emissivity, due to such factors as grain growth
through sticking and the formation of icy mantles.  As a result of
these processes, the dust emissivity in regions with higher column
density and lower dust temperature is seen to be higher.  This effect
has been observed in many environments (e.g. Kiss et al. 2006; del
Burgo \& Laureijs 2005; Bernard et al. 1999), where it is shown that
the FIR emissivity ($\alpha$ in Eq.~\ref{DUSTFLUX} or $X_\lambda$ in
Eq.~\ref{AVEQ}) anti-correlates with dust temperature and varies by a
factor of a few in interstellar clouds similar to Perseus.  In
Fig.~\ref{TAUAVTEMP} we show the FIR dust emissivity (expressed as the
ratio of the 100 \micron\ optical depth and NIR-derived extinction)
plotted against the color temperature of the dust.  We also find this
dependence of the emissivity on temperature and column density, which
suggests that the linear fit between $\tau_{FIR}$ and A$_V$ (see
Eq.~\ref{AVEQ} and Fig.~\ref{TAUAVTEMP}) does not fully describe the
relationship between dust emission and absorption.

It is important to note that temperature fluctuations along the line
of sight can give the appearance of a temperature-dependent emissivity
when the dust is assumed to be isothermal, even if the true dust
emissivity is constant.  Because the line-of-sight averaged dust
temperature is emission weighted, the dust temperature is
overestimated and the far-infrared optical depth is underestimated.
This effect is larger along lines of sight with greater temperature
variation (i.e., those passing through denser regions), which are
preferentially colder than the more diffuse (and nearly isothermal)
portions of the molecular cloud.  As a consequence, the derived ratio
of the far-infrared optical depth to extinction will be underestimated
in the colder, denser regions of the molecular cloud.  However, the
observed temperature-dependent dust emissivity varies in the {\it
  opposite direction} of the correlation created by line-of-sight
temperature fluctuations.  The observed dust emissivity is {\it
  larger} at low temperatures, supporting the claim that the true dust
emissivity is variable.

\begin{figure}
\epsscale{1.1}
\plotone{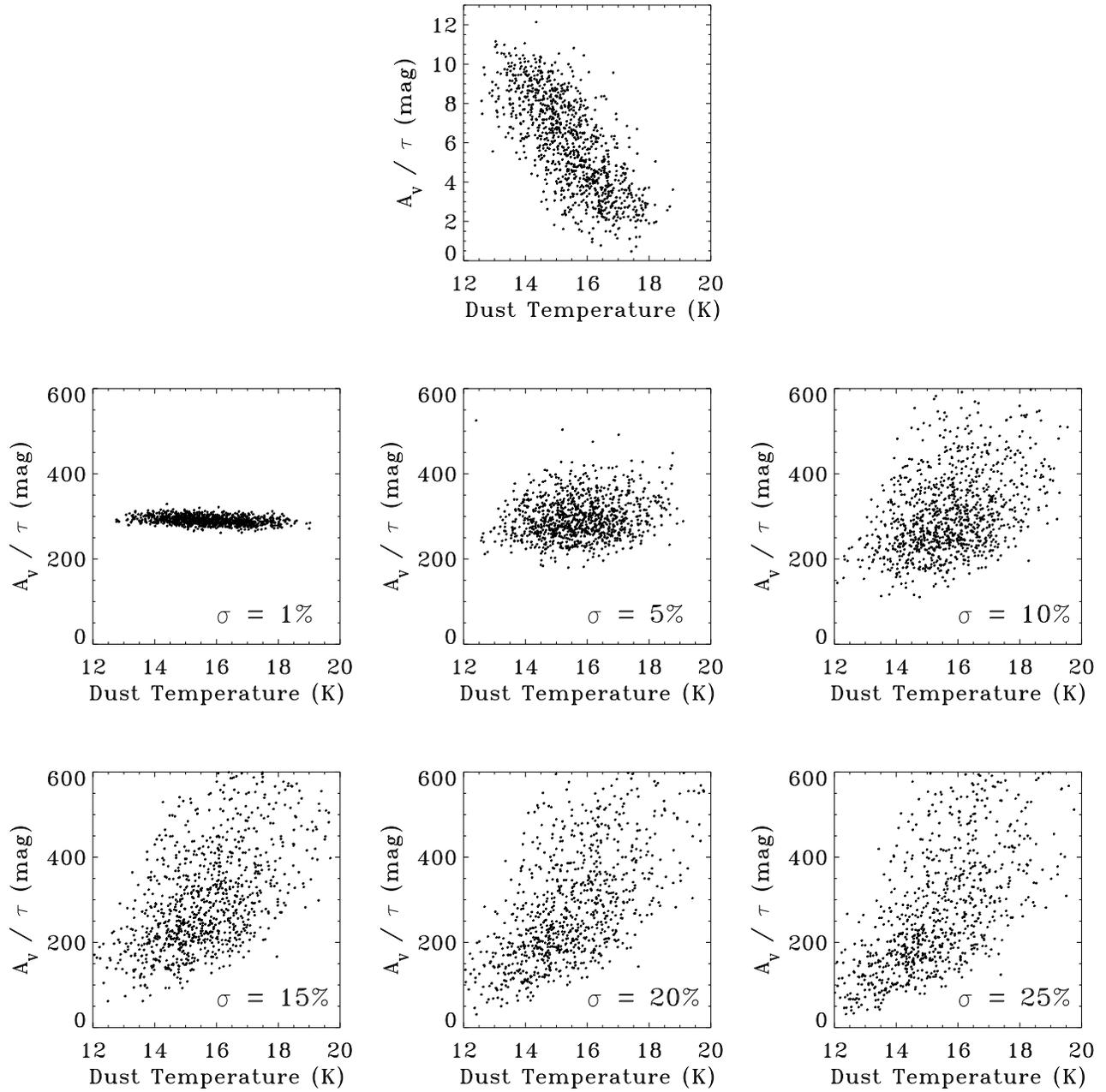}
\caption{({\it Top}) A model temperature and column density
distribution, which will be used to generate synthetic emission maps
and test the effects of noise on the derived dust emissivity. ({\it
Middle and Bottom}) The dust emissivity, expressed as the ratio of the
100 \micron\ optical depth and the column density, plotted against
dust temperature.  The optical depth and dust temperature are derived
from synethic flux maps at 100 and 160 \micron, with relative noise
levels ranging from 1\% to 25\%.  The ``true'' flux is calculated from
the column density and temperature distibution in the top panel, and
is then adjusted by Gaussian random noise at the level shown in each
middle and lower panel. \label{NOISEFIG}}
\end{figure}

Noise in the FIR emission maps can also give the appearance of a
variable dust emissivity.  If noise changes the ratio of the fluxes,
then the dust temperature will be either overestimated or
underestimated.  A high dust temperature will result in a low
FIR-derived optical depth, and a low dust temperature will result in a
high FIR optical depth.  To illustrate the effect of noise on the
derived dust emissivity, we analyze a simple model of a molecular
cloud with constant dust emissivity in which the temperature is
anti-correlated with column density and with both quantities are
within a typical range for molecular clouds.  We then derive the flux
at 100 and 160 \micron\ flux that would be emitted and add Gaussian
random noise to the maps.  We re-derive the dust temperature and 100
\micron\ optical depth from the noisy maps and show the dust
emissivity that would be estimated from the ``observed'' maps plotted
against the temperature.  The results are shown in
Fig.~\ref{NOISEFIG}.  The dust emissivity that we derive in Perseus
(see Fig.~\ref{AVSCATTER}) looks similar to what one would expect from
maps of dust that have a constant emissivity, but that have noise on
the level of $\sim$15\%.  However, the random noise in all of our maps
is much lower than this ($\lesssim$1\%), so we can rule out the
possibility that noise in the Perseus emission maps is tricking us
into thinking that the dust emissivity is variable.

To account for the variable dust emissivity in the determination of
the dust column density, we divide our data into the same ten equally
populated bins as we did in Section \ref{TD}, based on their
NIR-derived extinction and temperature.  Using Eq.~\ref{AVEQ}, we
derive the conversion ($X_\lambda$) separately for each bin and each
pair of emission maps.  The derived dust emissivity, as a function of
$T_d$, is plotted in Fig.~\ref{BGCONVERSION}.

\section{Results} \label{RESULTS}

\subsection{VSG Emission and Dust Emissivity vs. $T_d$ and $A_V$} \label{ENVIRONMENT}

As shown in Fig.~\ref{BGCONVERSION}, the 60 and 70 \micron\ emission
is dominated by the emission from VSG's.  The fractional contribution
of BG emission rises with increasing $T_d$ at both 60 and 70 \micron.
This is no surprise, given that warmer dust will have larger 60/100
and 70/160 \micron\ flux ratios.  At 70 \micron, the BG contribution
is a larger percentage of the total flux for $A_V > 2.6$ than $A_V <
2.6$, while at 60 \micron\ the difference is smaller and changes sign.
This difference may be caused by the 60 and 70 \micron\ passbands,
since IRAS is sensitive (sensitiviy $\ge$ 10\%) from 37 to 82
\micron\, while Spitzer is sensitive from 55 to 86 \micron.

The parameter used to derive the dust column density from the FIR
optical depth of the dust, as given by $X_\lambda$ in Eq.~\ref{AVEQ},
varies nearly identically with $T_d$ for all three pairs of emission
maps.  The evolution in $X_\lambda$ is primarily caused by the
variable emissivity of the dust (shown in Fig.~\ref{TAUAVTEMP}), but
also by line-of-sight temperature fluctuations and noise in the
emission maps.  Taking into account variations in the VSG contribution
to emission from dust with $\lambda < 100$ \micron\ and variations in
the emissivity of the BG's represents an increase in the complexity of
the analysis performed in SRGL.

\subsection{Emission vs Absorption-derived Column Density} \label{SCATTERS}

As discussed in SRGL, the point to point scatter between emission and
absorption-derived column density can be quite large (a few tens of
percent).  This scatter can be attributed to line-of-sight temperature
fluctions, non-constant dust emissivity and variations in the fraction
of emission coming from VSG's (though this last source of error should
not affect the column density derived from emission maps at or above
$\sim$100 \micron).

\begin{figure}
\epsscale{0.8}
\plotone{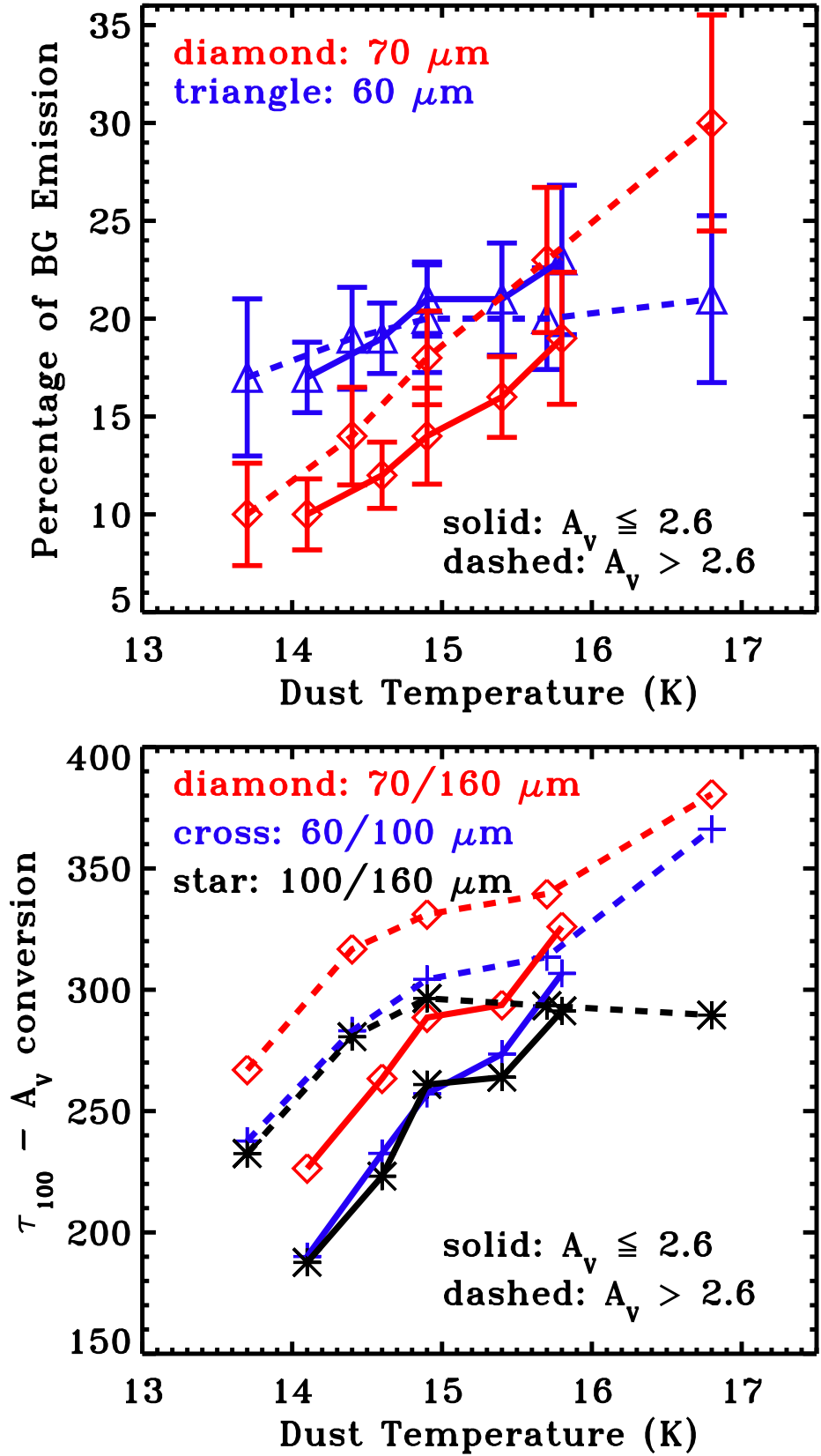}
\caption{({\it top}) The percentage of the observed 60 and 70 \micron\
flux emitted by BG's in thermal equilibrium as a function of dust
temperature.  The remaining portion is emitted by stochastically
heated VSG's.  The error bars show the standard deviation in each
bin. ({\it bottom}) The ratio of the observed extinction to the 100
\micron\ optical depth as a function of dust temperature.  For the
70/160 \micron\ pair of emission maps, we derive $\tau_{100}$ assuming
that $\beta = 2$ and scaling $\tau_{160}$ by $(100/160)^\beta$.  In
both the top and bottom plots there are $151 \pm 15$ independent
5\arcmin\ data values in each bin.
\label{BGCONVERSION}}
\end{figure}
\begin{figure}
\epsscale{1.0}
\plotone{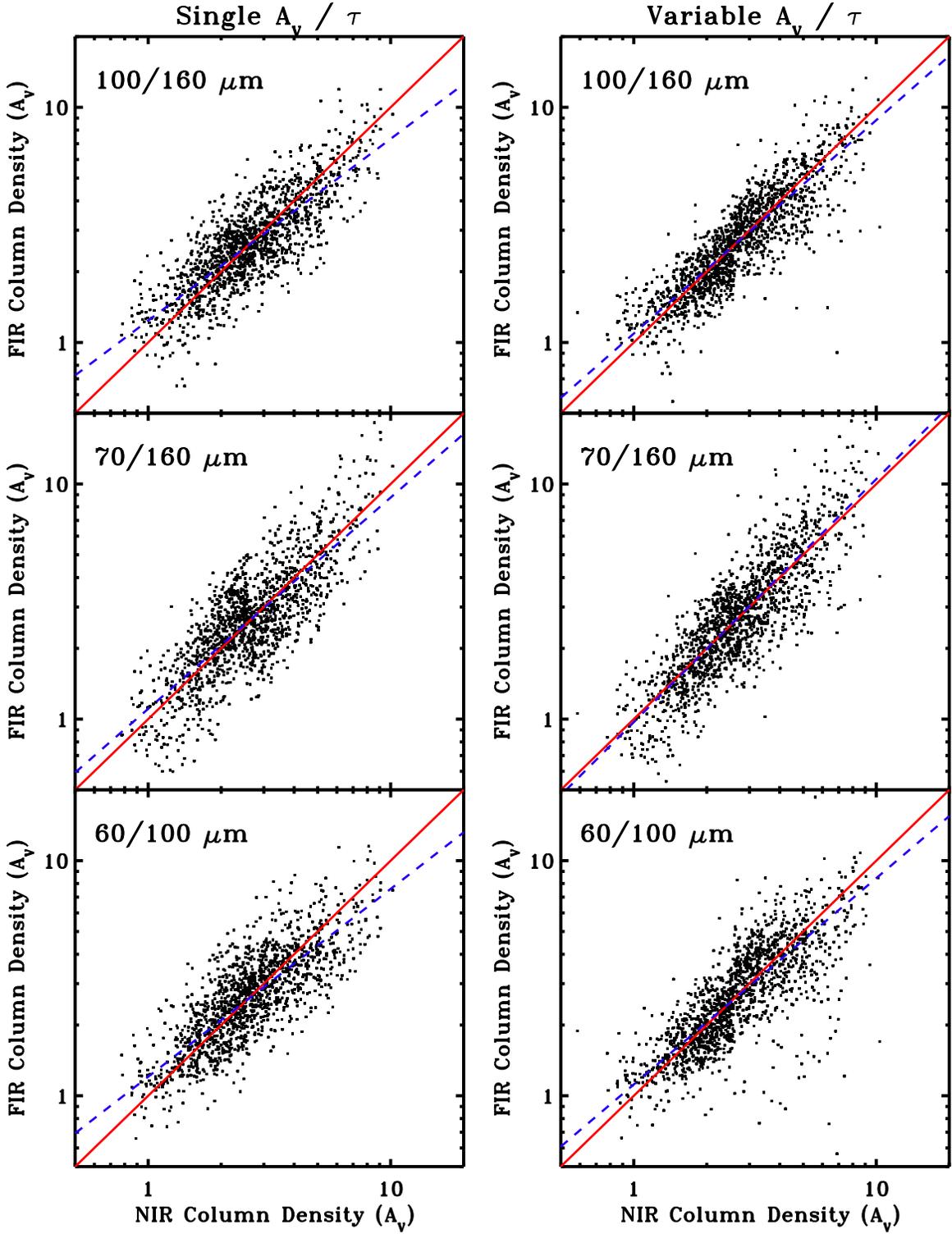}
\caption{({\it left}) The FIR-derived column density plotted against 
the NIR-derived column density.  Here we assume a single value for the
dust emissivity for each pair of emission maps.  ({\it right}) The
FIR-derived column density plotted against the NIR-derived column
density.  Here we assume a variable value for the dust emissivity for
each pair of emission maps, as plotted in Fig.~\ref{BGCONVERSION}.
\label{AVSCATTER}}
\end{figure}

In this paper we calculate the emission-derived column density in two
ways.  In the first method we assume that the dust emissivity is
constant in Perseus, and that $A_V$ scales with the FIR optical depth
(Eq.~\ref{AVEQ}).  Our second method uses a dust emissivity which
varies with $A_V$ and $T_d$, as shown in Fig.~\ref{BGCONVERSION}.  In
both cases we correct for variations in the relative VSG/BG
contribution to the 60 and 70 \micron\ emission, as a function of
$T_d$ and $A_V$, as described in Section \ref{TD}.  The values for the
median column density ratio, the scatter around that median, and the
slope of the best fit line between the emission vs absorption-derived
column density for Perseus are shown in Table \ref{COLUMNTAB} and the
corresponding plots are shown in Fig.~\ref{AVSCATTER}.

In SBG, we use a simplified model of emission from an externally
heated molecular cloud to study the effects of line-of-sight
temperature variations on column density estimates without having to
worry about the effects of VSG emission or variations in the dust
emissivity.  Using that model, we predict that the scatter and bias
should be largest for the 60/100 \micron\ pair of maps and smallest
for the 100/160 \micron\ pair.  However, in this paper, when we use a
single $\tau - A_V$ conversion, we find that the slope and scatter
between our emission and absorption-derived column densities are very
similar for all three pairs of emission maps, in conflict with SBG.
The amount of scatter in the NIR vs. 60/100 \micron-derived column
density is consistent with that predicted by the model in SBG. The
larger than expected scatter seen in the 70/160 is probably not caused
by an improper VSG correction, since we also find excess scatter in
the NIR vs. 100/160 \micron-derived column density, and VSG emission
should be insignificant in the 100 and 160 \micron\ maps.

When we divide the emission maps into ten subsets and derive the $\tau
- A_V$ conversion for each bin (shown in Fig.~\ref{BGCONVERSION}),
there is less bias in the emission-derived column density.  This is
shown by the fit between the FIR and NIR-derived column density in
Fig.~\ref{AVSCATTER}, which has a slope $\sim$1.  However, the scatter
in the relation (see Table \ref{COLUMNTAB}) is only significantly
improved for the 100/160 \micron\ derived column density compared to
the constant $\tau - A_V$ derivation.  This suggests that the emission
properties of dust are more complicated than assumed in this paper.
For instance, the emissivity (or spectral index) of the dust could be
density-dependent or there could be variations in grain properties as
a function of position in Perseus, and not merely as a function of
dust temperature and column density.

We conclude that the 100/160 \micron\ pair of maps is the best for
deriving dust temperature and column density because it is least
affected by the emission from VSG's and has the least NIR vs. FIR
column density scatter.  However, since the 70/160 pair of emission
maps has much better resolution, the column density and dust
temperature maps produced from it are also useful if one is willing to
accept the additional uncertainty of using maps presented at
$\sim40$\arcsec\ that have been calibrated on a $\sim5$\arcmin\ scale.

\subsection{Temperature and Column Density Maps} \label{MAPS}

\begin{figure}
\epsscale{1.1}
\plotone{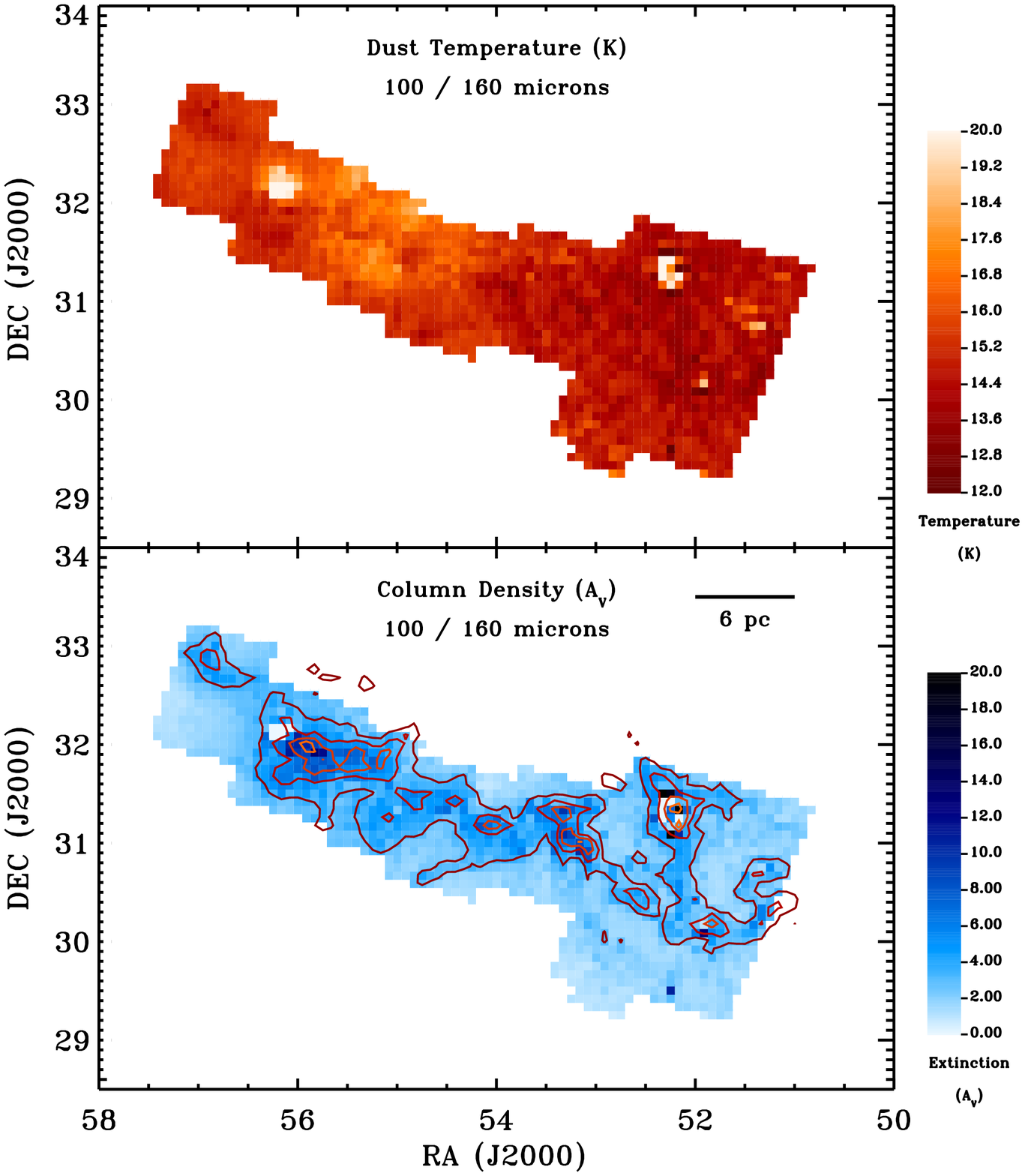}
\caption{({\it top}) The dust temperature map and ({\it bottom})
column density (expressed in terms of $V$-band extinction) derived
from the IRAS 100 and Spitzer 160 \micron\ flux density maps, at the
5\arcmin\ resolution of our NIR-derived extinction map.  The column
density is derived assuming a variable dust emissivity.  Red contours
in the bottom plot show the regions with extinction of $A_V = 3, 5, 7,
9$.
\label{MAPS160100}}
\end{figure}

\begin{figure}
\epsscale{1.1}
\plotone{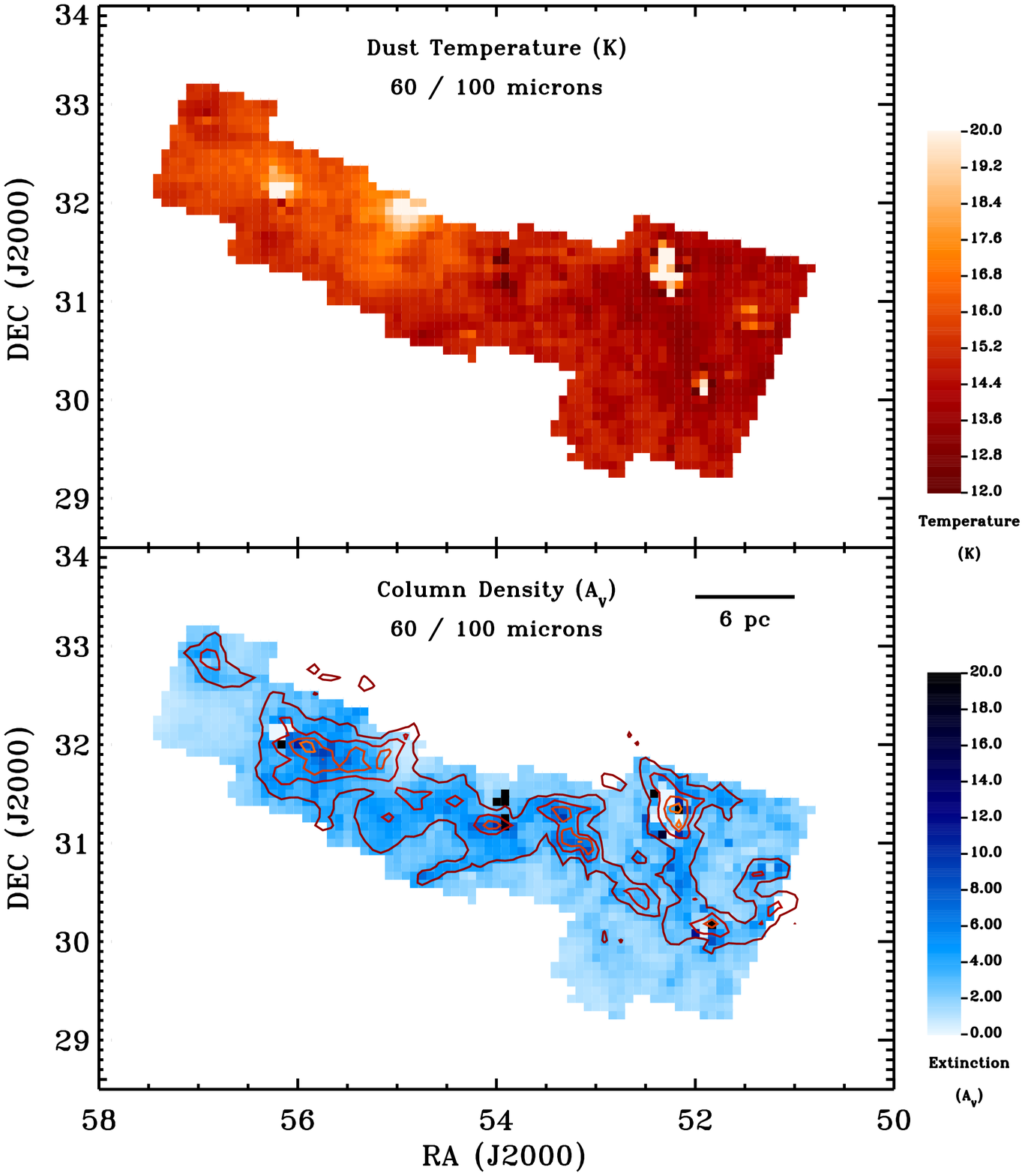}
\caption{({\it top}) The dust temperature map and ({\it bottom})
column density (expressed in terms of $V$-band extinction) derived
from the IRAS 60 and 100 \micron\ flux density maps, at the 5\arcmin\
resolution of our NIR-derived extinction map.  The column density is
derived assuming a variable dust emissivity.  Red contours in the
bottom plot show the regions with extinction of $A_V = 3, 5, 7,
9$. \label{MAPS10060}}
\end{figure}

\begin{figure}
\plotone{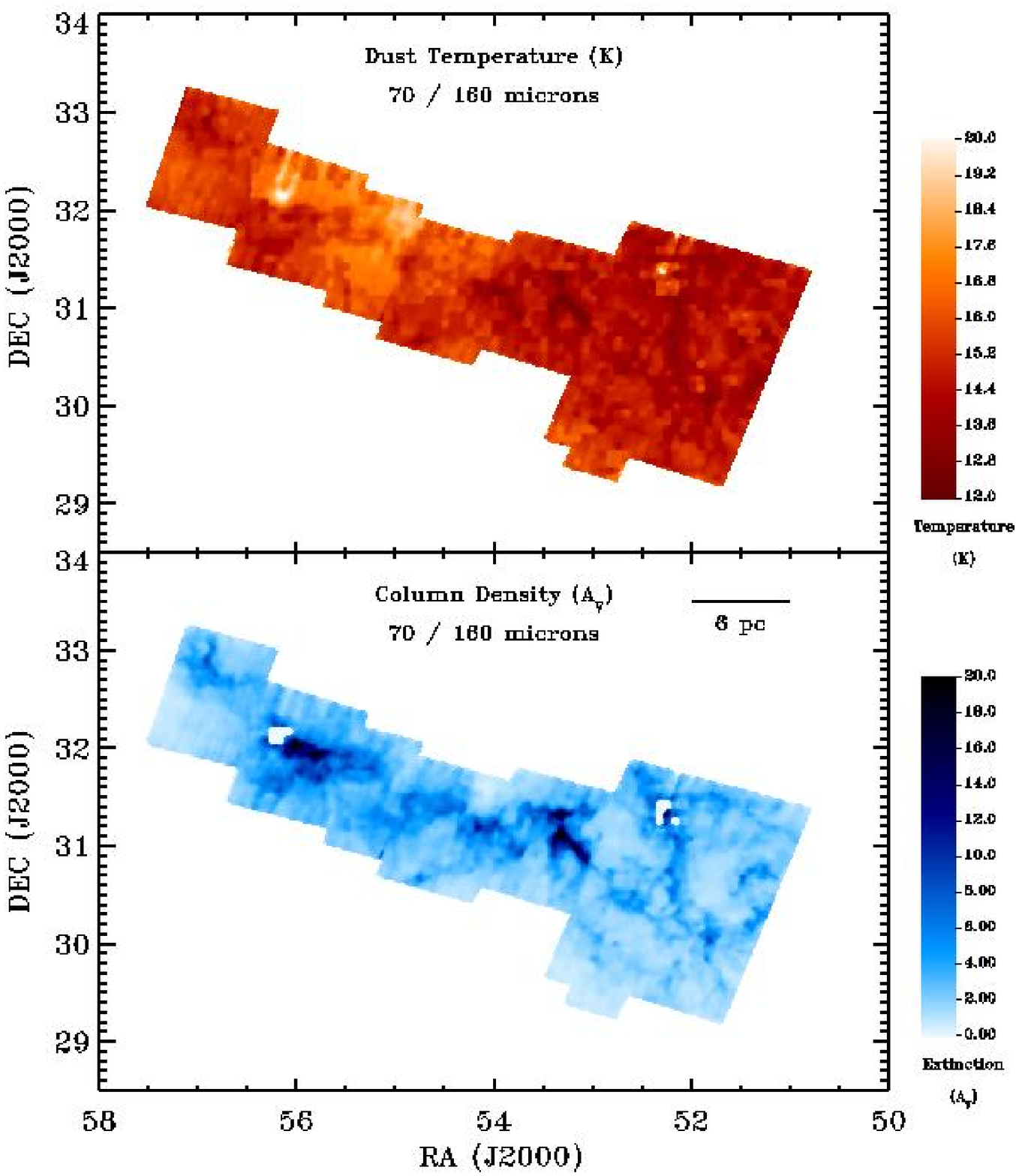}
\caption{({\it top}) The dust temperature map and ({\it bottom})
column density (expressed in terms of $V$-band extinction) derived
from the Spitzer 70 and 160 \micron\ flux density maps, at the
40\arcsec\ resolution of the 160 \micron\ map.  The column density is
derived assuming a variable dust emissivity.  Red contours in the
bottom plot show the regions with extinction of $A_V = 3, 5, 7,
9$.\label{MAPS16070HIGH}}
\end{figure}

Here we present dust temperature and column density maps derived from
the IRAS and Spitzer fluxes.  Figures \ref{MAPS160100},
\ref{MAPS10060} and \ref{MAPS16070} show the temperature and column
density maps derived from the IRAS and Spitzer far-infrared data and
are presented at $\sim$5\arcmin\ resolution (which is the resolution
of the NIR-derived extinction map and is approximately the resolution
of the IRIS 60 and 100 \micron\ maps).  Figure \ref{MAPS16070HIGH} is
derived from the Spitzer 70 and 160 \micron\ data, and is presented at
the 40\arcsec\ resolution of the 160 \micron\ map (the 70 \micron\
emission map was smoothed to the resolution of the 160 \micron\ map).
The column densities are determined by assuming a variable dust
emissivity, as described in Section \ref{VARDUST}.  Note that the VSG
subtraction and variable dust emissivity are calibrated at 5\arcmin\
resolution, so the calibration of the dust temperature and column
density using the Spitzer 70 and 160 \micron\ maps present at
40\arcsec\ resolution is more uncertain than the maps presented at
5\arcmin\ resolution.

\subsection{Temperature and Column Density Histograms} \label{HISTOGRAMS}

\begin{figure}
\epsscale{1.0}
\plotone{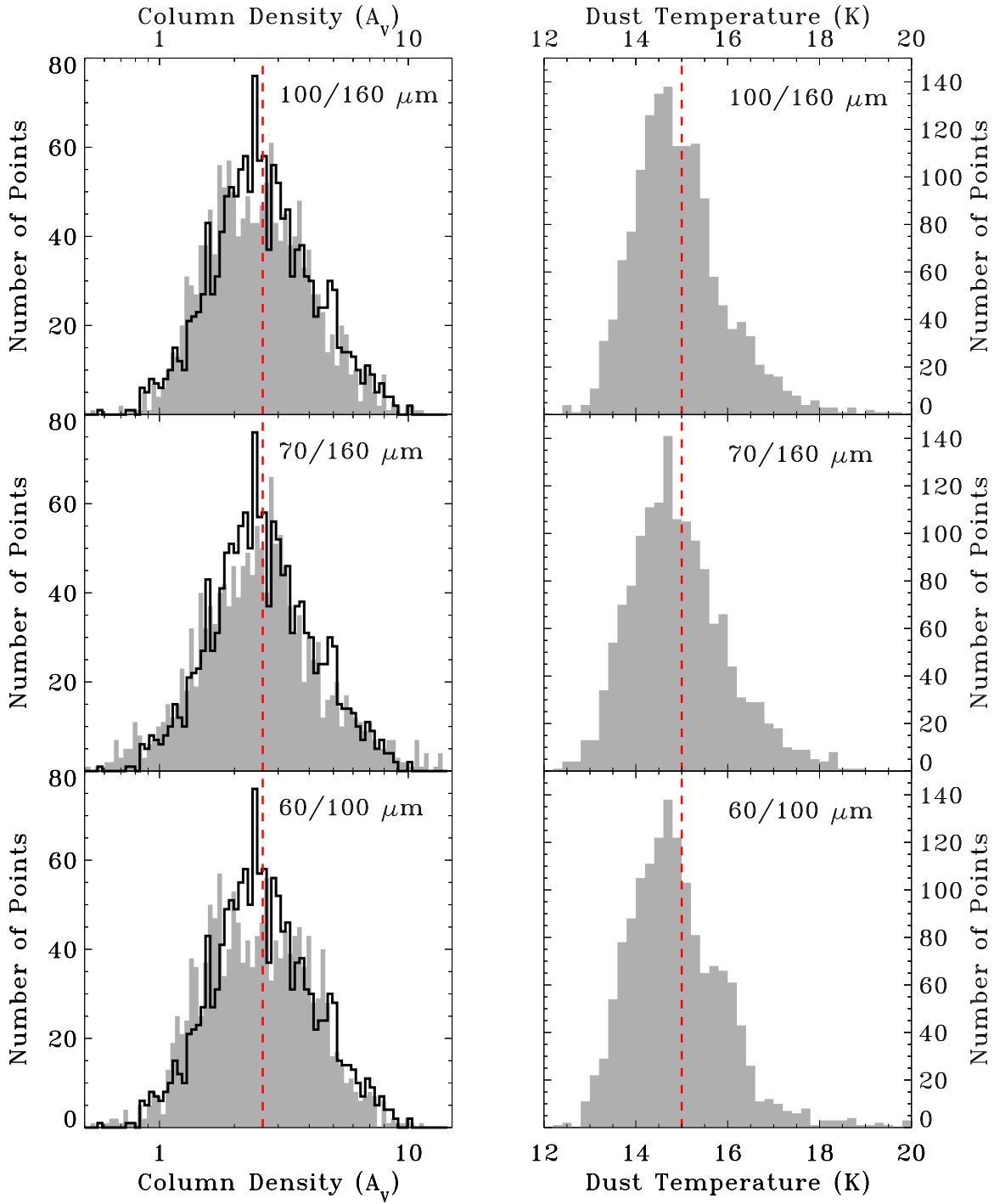}
\caption{Histograms of the dust color temperature and column density 
derived from the IRAS 60 and IRAS 100 \micron\ flux density maps ({\it
bottom}), Spitzer 70 and 160 \micron\ flux density maps ({\it middle})
and IRAS 100 and Spitzer 160 \micron\ flux density maps ({\it top}).
The column density is derived assuming a variable dust emissivity.
The dust temperature is derived assuming a variable fraction of 60 and
70 \micron\ flux emitted by BG's.  The filled histograms show the
column density and temperature derived from FIR emission, while the
open histograms show the column density derived from NIR absorption.
The vertical dashed lines show the median $A_V$ in the 2MASS map ({\it
left}) and the median dust temperature in the 100/160 \micron\ $T_d$
map ({\it right}). \label{PLOTHIST}}
\end{figure}

Figure \ref{PLOTHIST} shows histograms of the dust temperature and
column density distributions derived from from pairs of emission maps
at 60/100, 70/160 and 100/160 \micron.  The 60 and 70 \micron\ fluxes
have been scaled to remove the VSG emission (as explained in Section
\ref{TD}).  The median temperature derived for this region of the
Perseus molecular cloud by \citet{Schlegel98} is 16.5 K, whereas we
derive a median temperature of 15.0 K, so our derived temperature is
about 10\% lower.  The emission-derived column density is scaled to
match that derived from the NIR reddening of background stars.
Therefore, it is no surprise that the $A_V$ distributions in all three
panels of Fig.~\ref{PLOTHIST} are similar to that of the extinction
map.  

\section{Conclusion}

\begin{figure}
\epsscale{1.1}
\plotone{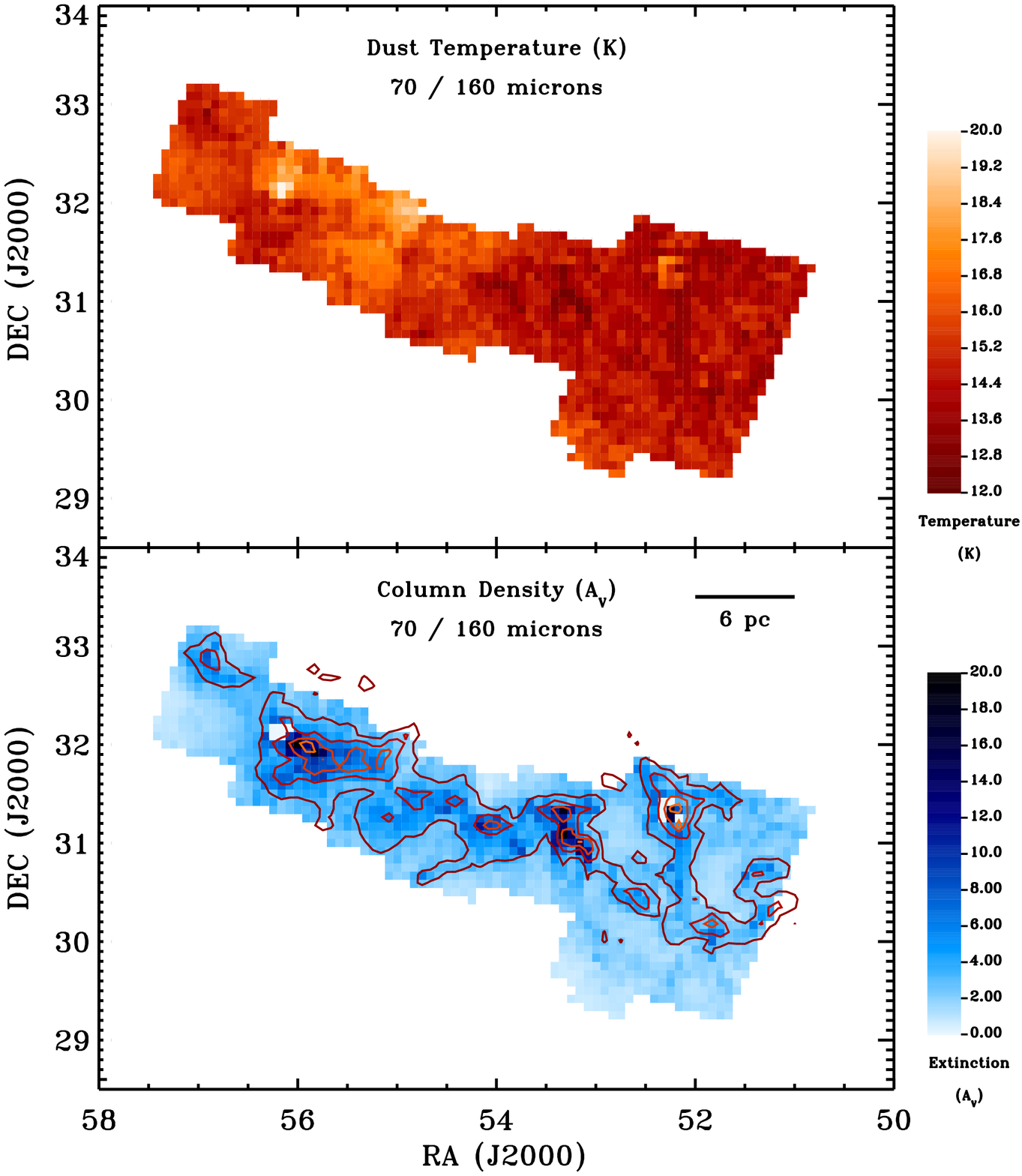}
\caption{({\it top}) The dust temperature map and ({\it bottom})
column density (expressed in terms of $V$-band extinction) derived
from the Spitzer 70 and 160 \micron\ flux density maps, at the
5\arcmin\ resolution of our NIR-derived extinction map.  The column
density is derived assuming a variable dust emissivity.  Red contours
in the bottom plot show the regions with extinction of $A_V = 3, 5, 7,
9$.\label{MAPS16070}}
\end{figure}

Using Spitzer 70 and 160 \micron\ maps released by the c2d Legacy
Project, we have created high-resolution dust temperature and column
density maps of the Perseus molecular cloud.  Although smaller in area
than similar maps presented in SRGL using IRAS and 2MASS data, the
maps in this paper (Fig.~\ref{MAPS16070}) are better calibrated to
account for variations in the dust emissivity and removal of VSG
emission.  Combining the Spitzer 160 \micron\ map with the IRAS 100
\micron\ map we show that the dust temperature in Perseus is
significantly colder (by $\sim$10\%) than reported in the dust
temperature map derived by \citet{Schlegel98}.  

We find that the dust emissivity varies with $A_V$ and $T_d$, in
agreement with the results of \citet{delBurgo05} and \citet{Kiss06}.
By fitting the absorption-derived column density to the FIR opacity in
each of ten subsets of the data, we demonstrate an improved method for
calculating the emission-derived dust column density.  The column
density derived from the 60/100 \micron\ pair of emission maps, when
compared with the NIR $A_V$, has slope and scatter consistent with
those expected from variations in temperature along the line of sight.
However, the scatter between the absorption-derived and 70/160 and
100/160 \micron\ emission-derived column density is larger than would
be expected from line-of-sight variations in the dust temperature
alone.  One possible explanation for the large scatter is that the 160
\micron\ map is more sensitive to the emission from cold dust grains
than are the 60 and 100 \micron\ maps (SBG), and the dust emissivity
varies more strongly at colder (T$_d < 15$ K) temperatures.  Although
the origin of the observed temperature dependence is uncertain, recent
theoretical models of amorphous dust grains by \citet{Meny07} and
\citet{Boudet05} also exhibit an anti-correlation between dust
temperature and emissivity.  Another possible cause for the higher
dust emissivity at lower temperatures is that the dust grains could be
accreting icy mantles in the dense and colder regions within the
molecular cloud.  Regardless of the reason, it is not surprising that
dust grains with a higher emissivity, given their ability to cool more
efficiently, will be found at lower equilibrium temperatures.

A temperature-dependent dust emissivity presents a challenge for those
attempting to derive dust temperatures and column densities from FIR
emission maps.  In SBG, we suggest that longer-wavelength ($\lambda >
100$ \micron) emission maps should be used to calculate the
line-of-sight dust properties because of their decreased sensitivity
to fluctuations in the dust temperature and the lack of confusing
emission from VSG's.  In this paper, we show that the dust emissivity
anti-correlates with dust temperature, and that the variability is
likely to increase at lower temperatures, an effect which is also seen
in other regions \citep[e.g.,][]{delBurgo05, Kiss06}.  This
complicates the interpretation of longer wavelength emission maps,
which are more sensitive to cold dust.  The deleterious effects of VSG
emission, temperature gradients, and variable dust emission properties
add complexity to the determination of the mass and temperature
distributions of dust in a molecular cloud.  Resolution of these
problems may require coverage from tens to hundreds of microns, finer
spectral resolution than the broad IRAS and Spitzer filters provide,
and comparison with dust emission models of ``realistic'' molecular
clouds.

Given the complexity of determining column density from dust emission,
one may wonder if using extinction might be a superior method.  For
instance, the extinction map of Perseus made with 2MASS data and the
NICER algorithm \citep{Ridge06, Lombardi01} is used as a template with
which we calibrate the emission-derived column density in this paper,
and we trust extinction-derived column density more than that derived
from dust emission in general \citep{Goodman08, Pineda08}.  The
resolution of extinction maps is limited by the number of background
tracers with known color, which provide 1\arcmin\ resolution for
clouds nearby the galactic center in projection \citep{Lombardi06},
but is coarser towards clouds further away from the galactic plane.
However, emission maps are limited only by the size of the detector
and also provide information on the dust temperature.

\acknowledgments

We would like to thank our anonymous referee for substantially
improving the scope and clarity of this paper.  Scott Schnee
acknowledges support from the Owens Valley Radio Observatory, which is
supported by the National Science Foundation through grant AST
05-40399.  Alyssa Goodman acknowledges support from the National
Science Foundation under Grant No. AST-0407172.

\clearpage

\begin{deluxetable}{lrrrrrrrrr} 
\tablewidth{0pt}
\tabletypesize{\scriptsize}
\tablecaption{Emission vs. Absorption-Derived Column Density 
              \label{COLUMNTAB}}
\tablehead{
 \colhead{Emission Maps}	     & \colhead{$\mu$\tablenotemark{a}}	   &
 \colhead{$\mu$\tablenotemark{b}}    & \colhead{$\sigma$\tablenotemark{c}} & 
 \colhead{$\sigma$\tablenotemark{d}} & \colhead{slope\tablenotemark{e}}    &
 \colhead{slope\tablenotemark{f}} \\
 \colhead{}                          & \colhead{constant}                  &
 \colhead{variable}                  & \colhead{constant}                  & 
 \colhead{variable}                  & \colhead{constant}                  &
 \colhead{variable}}
\startdata
 60/100 \micron  & 1.01 & 1.00 & 0.32 & 0.31 & 0.80 & 0.88 \\
 70/160 \micron  & 1.00 & 1.00 & 0.34 & 0.32 & 0.90 & 1.03 \\
 100/160 \micron & 1.00 & 1.00 & 0.30 & 0.25 & 0.77 & 0.91
\enddata
\tablenotetext{a}{Median ratio of $A_{V,emission}/A_{V,extinction}$ assuming a constant dust emissivity}
\tablenotetext{b}{Median ratio of $A_{V,emission}/A_{V,extinction}$ assuming a variable dust emissivity}
\tablenotetext{c}{1 $\sigma$ scatter in  $A_{V,emission}/A_{V,extinction}$ assuming a constant dust emissivity}
\tablenotetext{d}{1 $\sigma$ scatter in  $A_{V,emission}/A_{V,extinction}$ assuming a variable dust emissivity}
\tablenotetext{e}{Best fit slope from Figure \ref{AVSCATTER} (left)}
\tablenotetext{f}{Best fit slope from Figure \ref{AVSCATTER} (right)}
\end{deluxetable}

\clearpage

\end{document}